\documentclass[twocolumn,epjc3]{svjour3}
\usepackage[dvipsnames]{xcolor}

\usepackage{eso-pic}

\newcommand{\preprintA}{SI-HEP-2025-21}

\newcommand{\preprintfont}{\normalfont\rmfamily\footnotesize}
\newcommand{\preprintcolor}{black}

\newlength{\preprintTopMargin}    
\newlength{\preprintRightMargin}  
\makeatletter
\@ifundefined{shortstackgap}{\newlength{\shortstackgap}}{}
\makeatother

\newcommand{\PutPreprintFirstPage}{%
  \AddToShipoutPictureFG*{%
    \put(
      \LenToUnit{\dimexpr\paperwidth-\preprintRightMargin\relax},
      \LenToUnit{\dimexpr\paperheight-\preprintTopMargin\relax}
    ){%
      \makebox[0pt][r]{%
        \preprintfont \color{\preprintcolor}%
        \shortstack[r]{\preprintA\\\preprintB\\\preprintC}%
      }%
    }%
  }%
}

\journalname{Eur. Phys. J. C}
\usepackage{arydshln, cite, graphicx, hyperref, amsmath, newtxtext, newtxmath, relsize}
\hypersetup{colorlinks=true, linkcolor=ForestGreen, citecolor=ForestGreen, filecolor=ForestGreen, urlcolor=ForestGreen}

\begin{document}

\PutPreprintFirstPage

%
%\preprint{SI-HEP-2025-21}

%\begin{flushright}
%SI-HEP-2025-21\\
%\end{flushright}

\title{Extracting Higgs Self-Coupling Constraints through Triple Higgs Boson Production at Future Hadron Colliders}
\author{
  Benjamin Fuks\thanksref{i1,e1},
  Andreas Papaefstathiou\thanksref{i2,e2} and 
  Gilberto Tetlalmatzi-Xolocotzi\thanksref{i3,e3}
}

\institute{
  Laboratoire de Physique Th\'eorique et Hautes \'Energies (LPTHE), UMR 7589, Sorbonne Universit\'e et CNRS, 4 place Jussieu, 75252 Paris Cedex 05, France \label{i1} \and\ 
  Department of Physics, Kennesaw State University, 830 Polytechnic Lane, Marietta, GA 30060, USA \label{i2} \and\
  Theoretische Physik 1, Center for Particle Physics Siegen (CPPS), Universit\"at Siegen, Walter-Flex-Str. 3, 57068 Siegen, Germany\label{i3}
}

\thankstext{e1}{\href{mailto:fuks@lpthe.jussieu.fr}{fuks@lpthe.jussieu.fr}}
\thankstext{e2}{\href{apapaefs@kennesaw.edu}{apapaefs@kennesaw.edu}}
\thankstext{e3}{\href{gtx@physik.uni-siegen.de}{gtx@physik.uni-siegen.de}}

\date{\today}
\maketitle

\begin{abstract}
We present a systematic study of triple Higgs boson production at future high-energy hadron colliders, using the six-$b$-jet final state as a probe of the Higgs self-interactions. We conduct, under realistic detector smearing assumptions, both a traditional cut-based analysis, and a multivariate one using gradient boosting. The multivariate strategy is found to enhance sensitivity to beyond the Standard Model effects on the Higgs boson's self-couplings, while preserving large signal event yields, thus enabling more robust statistical inference. This allows us to assess the impact of detector effects, systematic uncertainties, background normalisation, as well as different truncation choices in an effective-field-theory description of the new physics effects possibly affecting the Higgs boson's self-interactions. Our results demonstrate that statistically-meaningful and perturbative-unitarity-compatible constraints on the trilinear and quartic Higgs boson self-couplings can be achieved, provided that  systematic uncertainties are controlled at the few-percent level. Finally, we extrapolate our results to various collider energies and luminosities, demonstrating in particular that an 85~TeV proton-proton collider performs comparably to a 100~TeV machine. Altogether, our findings therefore establish the six-$b$ channel as a viable probe of the Higgs self-interactions at most future hadron collider options currently being examined by the high-energy physics community.
\end{abstract}

\section{Introduction}\label{intro}
The discovery of a Higgs boson with a mass of about 125~GeV marked a milestone in particle physics~\cite{ATLAS:2012yve, CMS:2012qbp}, providing direct insight into the mechanism of electroweak symmetry breaking and the origin of fermion masses. Furthermore, subsequent studies of the Higgs boson's properties have shown remarkable agreement with the predictions of the Standard Model (SM), at least within the current experimental and theoretical uncertainties~\cite{ATLAS:2022vkf, CMS:2022dwd}. Nevertheless, establishing whether the observed state fully conforms to the Standard Model requires probing the shape of the Higgs potential itself, which, in turn, demands direct and independent determinations of the cubic and quartic Higgs self-couplings. More particularly, such measurements are not only central to validating the SM Higgs sector, but are also key to exploring scenarios of electroweak baryogenesis and the nature of the electroweak phase transition~\cite{Huang:2017jws, Ramsey-Musolf:2019lsf, Papaefstathiou:2020iag, Papaefstathiou:2021glr, Biermann:2024oyy, Karkout:2024ojx}. For these reasons, the determination of the Higgs self-couplings constitutes a major objective of the future high-energy physics programme~\cite{Contino:2016spe, FCC:2018byv, FCC:2018bvk, Cepeda:2019klc, Azzi:2019yne}.

The trilinear Higgs coupling represents the first step in this endeavour, and can be probed most directly in Higgs-boson pair production at hadron colliders~\cite{DiMicco:2019ngk}. In the SM, the corresponding cross section is sizeable, reaching about 38~fb at the LHC with a centre-of-mass energy of $\sqrt{s}=14$~TeV and rising to approximately 4.4~pb at $\sqrt{s}=100$~TeV~\cite{Borowka:2016ehy, Borowka:2016ypz, DeFlorian:2018eng, Baglio:2020ini, AH:2022elh}. Such a high rate, potentially further enhanced in scenarios where the destructive interference among the contributing diagrams is reduced, enable a broad exploration of di-Higgs boson final states. The most sensitive channels include the $4b$, $2b2\gamma$, $2b2\tau$ and $2\gamma2\tau$ signatures, which have been actively pursued in both ATLAS and CMS searches~\cite{ATLAS:2023elc, ATLAS:2023vdy, ATLAS:2024pov, ATLAS:2024lhu, ATLAS:2024ish, CMS:2018ipl, CMS:2020tkr, CMS:2022cpr}. In addition, the trilinear coupling leaves an imprint on single-Higgs production through higher-order loop corrections, providing indirect and complementary sensitivity~\cite{Degrassi:2016wml, Gorbahn:2016uoy, Bizon:2016wgr, Maltoni:2017ims, ATLAS:2022jtk, CMS:2024awa}.

The quartic Higgs self-coupling is the second essential ingredient for determining the shape of the Higgs potential, and it can similarly be probed either directly through triple-Higgs boson production, or indirectly via loop corrections to di-Higgs boson production processes. In the SM, however, the triple-Higgs boson cross section is extremely suppressed due to a strong destructive interference between the different one-loop contributions. At the LHC, the total rate is hence only of about 0.05~fb at $\sqrt{s}=14$~TeV~\cite{deFlorian:2019app}, making any observation entirely out of reach. At a future proton-proton collider with $\sqrt{s}=100$~TeV, the prospects are more promising, with the total cross section increasing to approximately 5.6~fb, although, in the presence of physics beyond the SM, this rate can be further enhanced through resonant contributions, or anomalous effects that modify the trilinear and/or quartic Higgs boson self-couplings. 

Triple-Higgs boson production leads to a broad variety of multi-particle final states, with the most extensively studied signatures so far include the $6b$ channel~\cite{Papaefstathiou:2019ofh, Papaefstathiou:2020lyp, Papaefstathiou:2023uum, Stylianou:2023tgg,Papaefstathiou:2025meh}, the $4b2\gamma$ final state~\cite{Papaefstathiou:2015paa, Fuks:2015hna, Chen:2015gva}, the $4b2\tau$ mode~\cite{Fuks:2017zkg, Stylianou:2023tgg}, the $2b4\tau$ channel~\cite{Stylianou:2023tgg, Dong:2025lkm}, as well as final states with multiple gauge bosons such as the $4W2b$ one~\cite{Kilian:2017nio}. Each of these provides different balances between signal rates and background rejection, with several channels showing an excellent discovery potential at 100~TeV. In addition, as for the case of the trilinear Higgs self-coupling, complementary indirect constraints can also be obtained from precision di-Higgs boson production measurements~\cite{Bizon:2018syu, Borowka:2018pxx, Bizon:2024juq}.

In light of these challenges, it is essential to assess the extent to which future high-energy hadron colliders can deliver meaningful constraints on the Higgs self-couplings through triple-Higgs boson production. In this work, we focus on the six $b$-jet final state, which, despite its complexity, offers a relatively large signal rate and robust sensitivity to both the trilinear and quartic Higgs couplings. Extending previous studies~\cite{Papaefstathiou:2019ofh, Papaefstathiou:2020lyp, Papaefstathiou:2023uum, Stylianou:2023tgg}, we employ two complementary analysis strategies, namely a traditional cut-based approach and a multivariate method based on gradient boosting~\cite{Cornell:2021gut}. Furthermore, in order to provide a realistic estimate of the expected sensitivity of future machines, we study different detector-smearing scenarios, and explore the impact of systematic uncertainties on the dominant backgrounds. We also address theoretical uncertainties like those inherent to high-order corrections to the background processes, and examine the impact of the number of new physics insertions allowed at the level of the signal squared matrix element. Finally, we extrapolate our results to several proton-proton collider centre-of-mass energies and integrated luminosities, relevant for the future collider configurations recently discussed within the community. Our analysis hence provides a comprehensive up-to-date picture of the prospects for constraining the Higgs boson's self-interactions beyond the LHC.

The remainder of this article is organised as follows. In section~\ref{sec:theory}, we introduce the employed theoretical framework and describe how new physics effects can be modelled, either through anomalous couplings or within a Standard Model effective field theory expansion. Moreover, we also briefly discuss perturbativity constraints on the relevant parameters. Section~\ref{sec:pheno} presents our simulation and analysis setup: we detail the treatment of detector smearing effects, the modelling of background contributions, and the definition of both our cut-based and multivariate strategies, before outlining the procedure used to extract the sensitivity reach of future colliders. Our results are reported in section~\ref{sec:results}, and we provide a summary and outlook in section~\ref{sec:conclusions}.

\section{Triple-Higgs Boson Production at Hadron Colliders}\label{sec:theory}

In this work, we investigate the potential of future colliders to probe the strength of the Higgs boson self-interactions. To this end, we employ a phenomenological Lagrangian containing the terms
\begin{equation}\label{eq:LgghhPheno}
  \mathcal{L} \supset - \frac{ m_h^2 } { 2 v } \Big( 1 + c_3 \Big) h^3 - \frac{ m_h^2 } { 8 v^2 }  \Big( 1  + d_4 \Big) h^4\;,
\end{equation}
where $m_h$ is the Higgs boson mass, $v$ the vacuum expectation value of the Higgs field, and where $c_3$ and $d_4$ parametrise anomalous, non-standard, contributions to the Higgs self-couplings. Such deviations may arise within effective field theories (EFTs) such as the Standard Model Effective Field Theory (SMEFT), in which the explicit ultraviolet completion of the Standard Model (SM) is replaced by an infinite set of higher-dimensional operators involving SM fields and respecting the SM symmetries~\cite{Weinberg:1978kz, Leung:1984ni, Buchmuller:1985jz, Grzadkowski:2010es, Elias-Miro:2013mua, Degrande:2020evl, Isidori:2023pyp}, or within the electroweak chiral Lagrangian with a light Higgs boson~\cite{Feruglio:1992wf, Bagger:1993zf, Koulovassilopoulos:1993pw, Burgess:1999ha, Wang:2006im, Grinstein:2007iv, Alonso:2012px, Buchalla:2013eza, Buchalla:2012qq, Alloul:2013naa, Buchalla:2013rka, Buchalla:2017jlu}. In the SMEFT, the Higgs field is assumed to be a complex doublet $H$ and the lowest-order contributions to the Higgs self-couplings arise from dimension-six ($D=6$) operators. However, if only $D=6$ operators are considered, the triple and quartic self-couplings are fully correlated, like for instance when only an operator of the form $\mathcal{O}_6 \sim c_6 |H|^6 / \Lambda^2$ with coefficient $c_6$ and new physics scale $\Lambda$ is included. This indeed leads to $c_3 = c_6$ and $d_4 = 6c_6$, as shown for instance in~\cite{Goertz:2014qta}. This correlation can be relaxed by introducing $D=8$ operators that would become relevant if new physics states lie close to the electroweak scale, such as $\mathcal{O}_8 \sim c_8 |H|^8 / \Lambda^4$ with coefficient $c_8$. In contrast, within the electroweak chiral Lagrangian the Higgs boson is treated as a singlet $h$, and no such correlations exist a priori: $c_3$ and $d_4$ are independent from the outset.

When calculating cross sections for processes affected by EFT contributions, a central issue is how many EFT operator insertions to retain at the matrix-element level, and how to truncate the EFT expansion at the squared amplitude level. At present, there is no general consensus on the appropriate truncation procedure~\cite{Degrande:2016dqg, Brivio:2022pyi, Alasfar:2023xpc}, and this ambiguity is particularly relevant for triple Higgs boson production since the three-body final state allows for two insertions of the triple self-coupling in a ``double-Higgstrahlung''-type topology. In addition, constraints on the quartic self-coupling are strongly entangled with large modifications of the triple self-coupling (see for instance~\cite{Papaefstathiou:2015paa, Fuks:2015hna, Chen:2015gva, Fuks:2017zkg, Kilian:2017nio, Papaefstathiou:2019ofh, Stylianou:2023tgg, Abouabid:2024gms, Dong:2025lkm}). To assess the numerical impact of different EFT truncation schemes, we consider four possibilities differing in how contributions from the SM amplitude $\mathcal{M}_\mathrm{SM}$, diagrams with one EFT insertion $\mathcal{M}_\mathrm{1ins}$ and diagrams with two insertions $\mathcal{M}_\mathrm{2ins}$ are combined. For a linear truncation, only the interference between the SM and one-insertion amplitudes is retained,
\begin{equation}
  |\mathcal{M}|^2 = |\mathcal{M}_\mathrm{SM}|^2 + 2\,\mathrm{Re}(\mathcal{M}_\mathrm{SM}^* \mathcal{M}_\mathrm{1ins})\;,
\end{equation}
while in the case of a quadratic truncation, the square of the one-insertion diagrams is also included together with the interference between the SM and two-insertion contributions,
\begin{equation}
  |\mathcal{M}|^2 = |\mathcal{M}_\mathrm{SM} + \mathcal{M}_\mathrm{1ins}|^2 + 2\,\mathrm{Re}(\mathcal{M}_\mathrm{SM}^* \mathcal{M}_\mathrm{2ins})\;.
\end{equation}
On the other hand, in a cubic truncation scheme an additional interference term between the one-insertion and two-insertion diagrams is kept,
\begin{equation}\begin{split}
  |\mathcal{M}|^2 = &\ |\mathcal{M}_\mathrm{SM}+ \mathcal{M}_\mathrm{1ins}|^2 + 2\,\mathrm{Re}(\mathcal{M}_\mathrm{SM}^* \mathcal{M}_\mathrm{2ins}) \\ &\ + 2\,\mathrm{Re}(\mathcal{M}_\mathrm{1ins}^* \mathcal{M}_\mathrm{2ins})\;,
\end{split}\end{equation}
whilst without any truncation the squared amplitude is simply taken as the modulus squared of the full sum of contributions,
\begin{equation}
  |\mathcal{M}|^2 = |\mathcal{M}_\mathrm{SM} + \mathcal{M}_\mathrm{1ins} + \mathcal{M}_\mathrm{2ins}|^2\;.
\end{equation}

\begin{figure}
    \centering
    \includegraphics[width=0.95\linewidth]{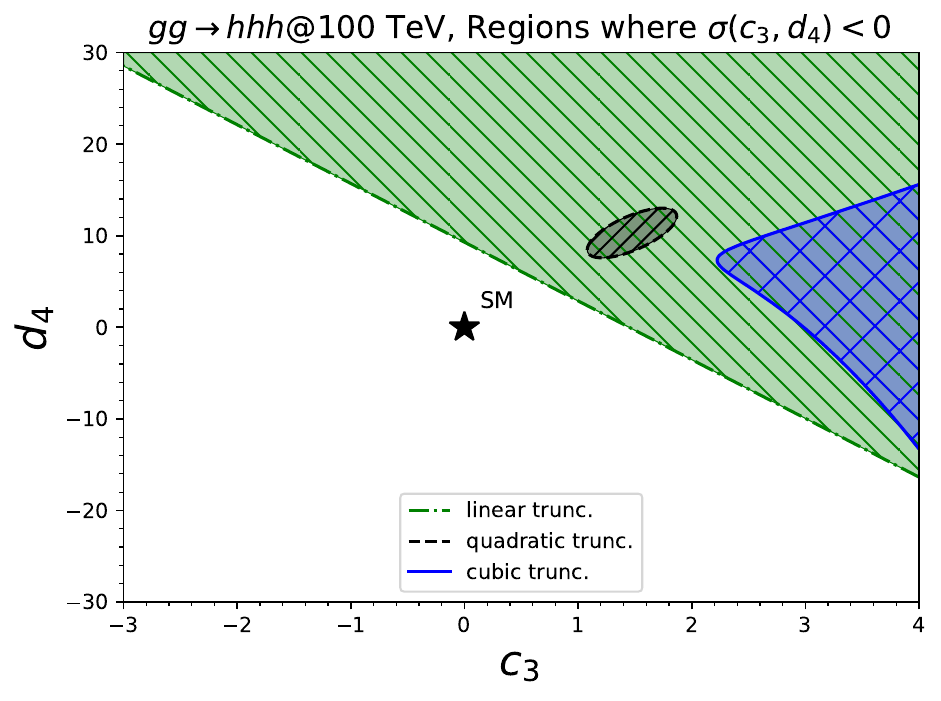}
    \caption{Regions of the $(c_3,d_4)$ parameter space where the $gg \to hhh$ cross section becomes negative under different truncation schemes: linear (green), quadratic (black) and cubic (blue). The non-truncated case remains positive definite for all values of $c_3$ and $d_4$.}
    \label{fig:negative}
\end{figure}

We now apply these schemes to triple-Higgs production in proton-proton collisions at a centre-of-mass of 100~TeV. The total production cross section $\sigma_{hhh}$ is found to depend on the $c_3$ and $d_4$ coefficients of the phenomenological Lagrangian terms shown in eq.~\eqref{eq:LgghhPheno} after normalisation to the SM prediction $\sigma_\mathrm{SM}$ as
\begin{equation}\label{eq:sigmac3d4}\begin{split}
\frac{\sigma_{hhh}}{\sigma_\mathrm{SM}} =&\ 1 - 0.1076 d_4 - 0.6899 c_3\\
  &\ + 0.01559  d_4^2 - 0.1447  c_3 d_4+ 0.7390 c_3^2\\
&\ +  0.04065 c_3^2 d_4 -  0.2077  c_3^3  \\
 &\ + 0.0308 c_3^4  \;.
\end{split}\end{equation}
In this expression, the first line corresponds to linear truncation, the second to quadratic truncation, the third to cubic truncation, and the last term to the quartic contribution present only if no truncation is imposed. The numerical coefficients were obtained with the \texttt{MadGraph5\_aMC\@NLO} (\texttt{MG5\_aMC}) event generator~\cite{Alwall:2014hca, Hirschi:2015iia}, the custom UFO~\cite{Degrande:2011ua, Darme:2023jdn} model developed in \cite{Papaefstathiou:2023uum} (see also section~\ref{sec:sim} for details), and after convoluting the matrix elements with the \texttt{NNPDF23\_lo\_as\_0130\_qed} parton density set~\cite{Ball:2012cx}. Furthermore, the fit of eq.~\eqref{eq:sigmac3d4} is found statistically consistent with previous results~\cite{Papaefstathiou:2015paa, Papaefstathiou:2023uum}, as well as with the analytic calculations of~\cite{Campbell:2025llb}. We emphasise that only the non-truncated cross section remains positive definite for arbitrary values of $c_3$ and $d_4$. In contrast, truncated descriptions can lead to unphysical regions of the parameter space where the predicted cross section becomes negative. This is illustrated in figure~\ref{fig:negative}, where we show that the rate for the process $gg \rightarrow hhh$ turns negative in distinct regions depending on the truncation scheme. The linear truncation is the most pathological case, and because the projected experimental bounds at future colliders fully cover the corresponding negative region (shown in green in the figure), we do not consider it further. The quadratic truncation scheme also becomes unreliable near $c_3 \sim 1$, so results approaching this region are included only for completeness. Nevertheless, combining triple-Higgs constraints in the $(c_3, d_4)$ plane with independent measurements of $c_3$ from other processes is still reliable, as $c_3$ is expected to be constrained at the percent level at future colliders~\cite{deBlas:2019rxi}.

When deriving the expected experimental constraints on the Higgs boson self-couplings, it is also instructive to contrast them with the limits imposed by tree-level perturbative unitarity~\cite{DiLuzio:2017tfn, Liu:2018peg, Stylianou:2023tgg}. To calculate these unitarity bounds, we follow the procedure of~\cite{Stylianou:2023tgg} which employs the Jacobi-Wick expansion. Considering $hh \to hh$ scattering, probability conservation implies that the perturbative expansion must satisfy the optical theorem, leading to an upper bound on the zeroth partial wave amplitude, $a_{ii}^0$, as follows:
\begin{equation}
    \left| \mathrm{Re} (a_{ii}^0) \right| \leq \frac{1}{2}\;,
\end{equation}
where
\begin{equation}\begin{split}
  a_{ii}^0 =&\ -\frac{ 3 m_h^2 \sqrt{s^2 - 4 m_h^2 s}}{32 \pi s \Big(s - m_h^2 v^2\Big)}\  \Bigg[ 
         (1 + d_4)(s- m_h^2)\\
    &\qquad + 3 (1 + c_3)^2 m_h^2 \\
    &\qquad - \frac{6 (1 + c_3)^2 m_h^2(s-m_h^2)}{s-4m_h^2} \log\left(\frac{s}{m_h^2} -3\right)\Bigg]\;,
\end{split}\end{equation}
Using this expression, we scan over the centre-of-mass energy $\sqrt{s} \in [200, 5000]$~GeV in the $(c_3,d_4)$ plane, and we require that perturbative unitarity is preserved at all scales to determine the viable regions of the parameter space. Notably, constraints on the triple self-coupling are dominated by low-energy scattering, whereas bounds on the quartic self-coupling arise from the high-energy regime, as already pointed out in~\cite{DiLuzio:2017tfn}.\footnote{For unitarity bounds derived using $n\rightarrow m$ amplitudes for longitudinal gauge bosons, see refs.~\cite{Chang:2019vez,Abu-Ajamieh:2020yqi}.}

\section{Simulation and Analysis Framework}\label{sec:pheno}
In this section, we describe the simulation toolchain and analysis framework used to study triple-Higgs boson production at future colliders. We begin by discussing, in section~\ref{sec:sim}, signal and background event generation, with a particular emphasis on detector simulation and its impact on jet reconstruction. Subsequently, in section~\ref{sec:analysis}, we present our analysis strategy, including a traditional cut-based approach, as well as a multivariate analysis using \textsc{XGBoost} for gradient boosting. 

\subsection{Event Generation and Detector Simulation}\label{sec:sim}
Signal events are generated within the \texttt{MG5\_aMC} platform, using the \texttt{heft\_loop\_sm} UFO model of~\cite{Papaefstathiou:2023uum} to incorporate anomalous triple and quartic Higgs self-couplings as defined in the Lagrangian of eq.~\eqref{eq:LgghhPheno}. More specifically, event generation is performed by typing, in the \texttt{MG5\_aMC} command line interface, the command
\begin{verbatim}
 generate g g > h h h [noborn=MHEFT QCD] \
   MHEFT^ 2<=X
\end{verbatim}
where the parameter \texttt{X} specifies the truncation scheme: \texttt{1} for linear, \texttt{2} for quadratic, \texttt{3} for cubic and \texttt{4} for no truncation.  

In our analysis, we focus on the $6b$ final state arising from the decays of the triple Higgs boson system, which requires particular care in the simulation of background processes involving $b$-quarks. More specifically, when the $b$-quarks originate from pure QCD production rather than from $Z$ or Higgs decays, we also use \texttt{MG5\_aMC} for event generation, but this time with generation-level cuts on the $b$ quarks' transverse momentum ($p_{T,b}$), pseudo-rapidity ($\eta_b$) and angular separation ($\Delta R(b,b)$),
\begin{equation}\label{eq:cuts}
  p_{T,b} > 30~\mathrm{GeV}\;, \quad |\eta_b| < 5.0 \quad\text{and}\quad \Delta R_{b,b} > 0.2\;.
\end{equation}
In line with previous studies~\cite{Papaefstathiou:2019ofh, Papaefstathiou:2020lyp}, the process $pp \to b\bar{b}b\bar{b}b\bar{b}$ is particularly demanding, and we therefore rely on the ``gridpack'' option of \texttt{MG5\_aMC} to obtain a statistically significant sample efficiently.\footnote{Event generation was performed on the OMNI cluster at the University of Siegen.} The resulting generation-level cross sections are summarised in table~\ref{table:processes}, whereas the impact of additional and reducible backgrounds is discussed below. The leading-order SM cross section was calculated to be $\sim$2.88~fb at 100 TeV. We note that at present, higher-order corrections for the inclusive signal cross section have only been calculated using the ``$\mathrm{FT}_\mathrm{approx}$'' approach at next-to-leading order~\cite{Maltoni:2014eza}, or in the (Born-improved) heavy-top mass limit at next-to-next-to-leading order~\cite{deFlorian:2019app}. In the latter, the approximate cross section was found to be $\sim5.6$~fb at 100 TeV, in line with a $k$-factor of 2.0 that we assume here. Furthermore, there exists no Monte Carlo event generator that can generate the signal processes with matching or merging of higher orders to the parton shower. Such an event generator would be feasible, even with approximate higher orders, e.g.\ as was done for Higgs boson pair production in ref.~\cite{Maierhofer:2013sha}, and we leave this for future work. 

Following hard-scattering signal and background event generation with \texttt{MG5\_aMC}, showering, hadronisation and the simulation of the underlying event are performed with the general-purpose event generator \texttt{HERWIG 7}~\cite{Bahr:2008pv, Gieseke:2011na, Arnold:2012fq, Bellm:2013hwb, Bellm:2015jjp, Bellm:2017bvx, Bellm:2019zci, Bewick:2023tfi}. Event analysis is subsequently carried out using the \texttt{HwSim} framework add-on~\cite{hwsim}.

\begin{table*}[t!]\renewcommand{\arraystretch}{1.5}
\begin{tabular*}{\textwidth}{@{\extracolsep{\fill}}l r|r r|r r@{}}
 \multicolumn{6}{c}{\bf {Signal versus Irreducible Backgrounds}}\\[.1cm]
Process & \multicolumn{1}{c}{$\sigma_\mathrm{GEN} \times k_\mathrm{fac} \times \mathrm{BR} \times \mathcal{P}(6b)$  [fb]} 
        & \multicolumn{1}{c}{$\epsilon_\mathrm{cuts}^\mathrm{CMS}$ } 
        & \multicolumn{1}{c}{$N^{\mathrm{cuts}}_{20~\mathrm{ab}^{-1}}$ } 
        & \multicolumn{1}{c}{$\epsilon_\mathrm{\texttt{XGBoost}}^\mathrm{CMS}$ } 
        & \multicolumn{1}{c}{$N^{\mathrm{\texttt{XGBoost}}}_{20~\mathrm{ab}^{-1}}$ } \\[.2cm]
\hline
$hhh$ (SM)                                      & $4.29 \times 10^{-1}$    & $6.22 \times 10^{-3}$ & $53.2$  
                                               & $4.39 \times 10^{-2}$   & $377$  \\\hline
QCD $(b\bar{b})(b\bar{b})(b\bar{b})$            & $1.07 \times 10^{4}$     & $1.84 \times 10^{-5}$ & $3.92 \times 10^{3}$ 
                                               & $1.37 \times 10^{-4}$   & $2.92 \times 10^{4}$ \\
$q\bar{q} \rightarrow Z(b\bar{b})(b\bar{b})$    & $3.61 \times 10^{2}$     & $2.00 \times 10^{-5}$ & $144$ 
                                               & $1.36 \times 10^{-4}$   & $985$ \\
$q\bar{q} \rightarrow ZZb\bar{b}$               & $1.14 \times 10^{1}$     & $5.00 \times 10^{-6}$ & $\mathcal{O}(1)$ 
                                               & $2.17 \times 10^{-4}$   & $49$ \\
$q\bar{q} \rightarrow ZZZ$                      & $1.80 \times 10^{-1}$    & $5.00 \times 10^{-6}$ & $\ll 1$ 
                                               & $1.77 \times 10^{-4}$   & $\mathcal{O}(1)$ \\
$q\bar{q} \rightarrow hZb\bar{b}$               & $2.04$                   & $5.50 \times 10^{-5}$ & $\mathcal{O}(1)$ 
                                               & $1.22 \times 10^{-3}$   & $50$ \\
$q\bar{q} \rightarrow hZZ$                      & $1.48 \times 10^{-1}$    & $4.50 \times 10^{-5}$ & $\ll 1$ 
                                               & $1.58 \times 10^{-3}$   & $5$ \\
$q\bar{q} \rightarrow hhZ$                      & $8.11 \times 10^{-2}$    & $3.05 \times 10^{-4}$ & $\mathcal{O}(1)$ 
                                               & $7.32 \times 10^{-3}$   & $12$ \\
$q\bar{q} \rightarrow hh(b\bar{b})$             & $1.80 \times 10^{-2}$    & $1.20 \times 10^{-4}$ & $\ll 1$ 
                                               & $2.20 \times 10^{-3}$   & $\mathcal{O}(1)$ \\
$q\bar{q} \rightarrow h(b\bar{b})(b\bar{b})$    & $7.25 \times 10^{-1}$    & $5.50 \times 10^{-5}$ & $\mathcal{O}(1)$ 
                                               & $7.84 \times 10^{-4}$   & $11$ \\
$gg \rightarrow ZZZ$                            & $5.18 \times 10^{-3}$    & $\sim 10^{-5}$        & $\ll 1$ 
                                               & $1.73 \times 10^{-3}$   & $\mathcal{O}(1)$ \\
$gg \rightarrow hZZ$                            & $3.59 \times 10^{-2}$    & $1.70 \times 10^{-4}$ & $\ll 1$ 
                                               & $4.79 \times 10^{-3}$   & $3$ \\
$gg \rightarrow hhZ$                            & $6.41 \times 10^{-2}$    & $2.40 \times 10^{-4}$ & $\mathcal{O}(1)$ 
                                               & $1.02 \times 10^{-2}$   & $13$ \\
\hdashline
$\sum$ backgrounds & & & $ 4.1 \times 10^{3}$ & \multicolumn{2}{r}{$3.16 \times 10^{4}$} \\
\end{tabular*}
\caption{List of Standard Model signal (upper panel, thus with $c_3=d_4=0$, total cross section $\sim2.88$~fb) and background (lower panel) processes considered in the six-$b$-jet analysis carried out in this work. The second column reports the generation-level cross sections, including a $k$-factor of~2, all relevant branching ratios and a $b$-tagging performance of $\mathcal{P}(b \to b)=0.85$. The third and fifth columns show the analysis efficiencies when employing a ``CMS-like'' smearing and either a cut-based approach or the \texttt{XGBoost}-based approach, respectively, while columns four and six give the corresponding event yields at an integrated luminosity of $20~\mathrm{ab}^{-1}$ and $\sqrt{s}=100$~TeV after cuts. Small expected yields are quoted schematically as $\mathcal{O}(1)$ or $\ll 1$.\label{table:processes}
}
\end{table*}

\begin{figure}
    \centering
    \includegraphics[width=0.95\linewidth]{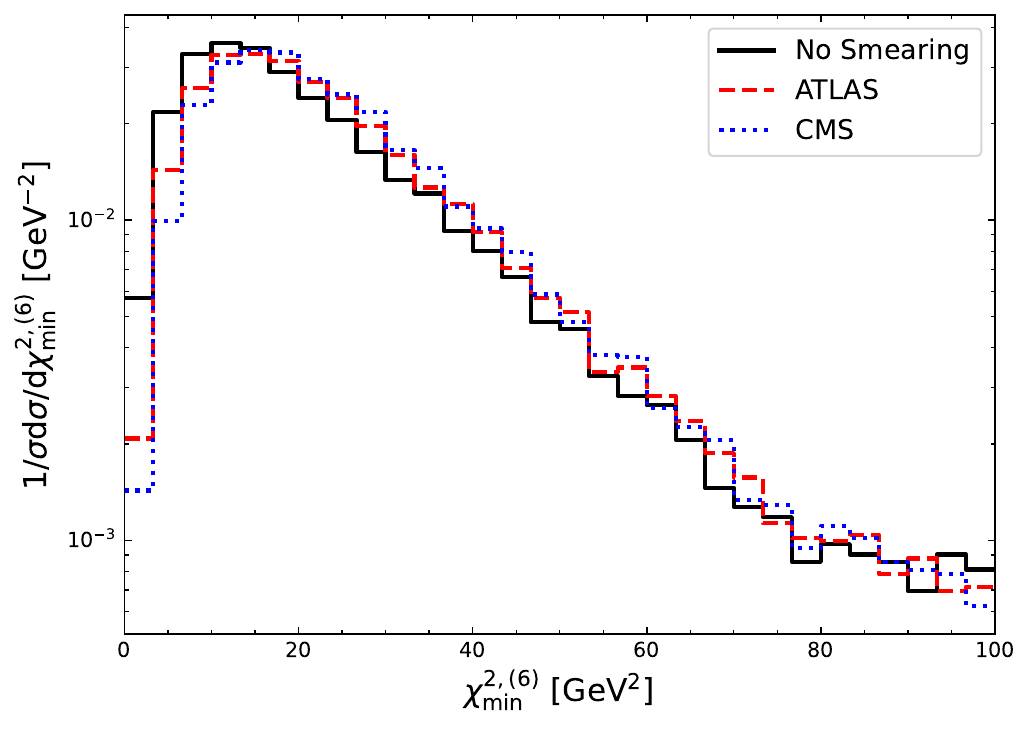}
    \caption{Impact of ATLAS-like and CMS-like jet smearing on the $\chi^{2,(6)}_\text{min}$ variable defined in section~\ref{sec:analysis}, in the case of SM triple Higgs boson production.}
    \label{fig:chisqsmear}
\end{figure}

To model the finite detector resolution, we implement jet smearing functions following existing ATLAS and CMS performance parametrisations as introduced in \textsc{MadAnalysis}~5~\cite{Conte:2012fm, Conte:2014zja, Conte:2018vmg, Araz:2020lnp}. The general idea is that the jet energy, after reconstruction with \texttt{FastJet}~\cite{Cacciari:2011ma} and the anti-$k_T$ algorithm~\cite{Cacciari:2008gp}, is subject to a Gaussian fluctuation around its true value,
\begin{equation}
  E_\text{smeared} = E + \Delta E\;, \qquad 
  \Delta E \sim \mathcal{N}(0,\,\sigma_E)\;,
\end{equation}
where $E$ is the true jet energy and $\sigma_E$ denotes the energy resolution. The latter is parametrised in terms of stochastic, noise and constant terms whose form depends on the jet energy and the pseudo-rapidity. For ATLAS-like and CMS-like scenarios we respectively adopt the parametrisations
\begin{equation}\begin{split}
  &\sigma^{\rm ATLAS}_{E} = \\
  &\ \begin{cases}
    \sqrt{(0.0302E)^2 \!+\! 0.5205^2 E \!+\! 1.59^2} & |\eta|\leq 1.7 \\[0.5ex]
    \sqrt{(0.05E)^2 \!+\! 0.706^2 E} & |\eta| \!\in\! [1.7, 3.2] \\[0.5ex]
    \sqrt{(0.0942E)^2 \!+\! E} & |\eta| \!\in\! [3.2, 4.9]
  \end{cases}\;, \\[.2cm]
  & \sigma^{\rm CMS}_{E} = \\
  &\ \begin{cases}
    \sqrt{(0.05E)^2 + 1.5^2 E} & |\eta|\leq 3.0 \\[0.5ex]
    \sqrt{(0.130E)^2 + 2.7^2 E} & 3.0<|\eta|\leq 5.0
  \end{cases}\;,
\end{split}\end{equation}
where $E$ and $\sigma_E$ are both expressed in GeV. We further assume that the angular coordinates of the jet four-momentum are unaffected by smearing so that the entire transformation reads
\begin{equation}
  (E, p_T, \eta, \phi) \;\to\;  (E_\text{smeared}, p_{T,\text{smeared}}, \eta, \phi)\;,
\end{equation}
with
\begin{equation}
 E_\text{smeared} = E + \Delta E \quad\text{and}\quad  p_{T,\text{smeared}} = \frac{E + \Delta E}{\cosh\eta}\;,
\end{equation}
and where $\Delta E$ is drawn from a Gaussian distribution of width $\sigma_E$ as specified above. As an illustration, it can be noted that for a central jet of energy $E=100$~GeV, the resolutions correspond to $\sigma_E^{\rm ATLAS} \simeq 6.2$~GeV and $\sigma_E^{\rm CMS} \simeq 15.8$~GeV for ATLAS-like and CMS-like jet smearing, respectively. Moreover, the relative energy resolution $\sigma_E/E$ improves with increasing jet energy due to the adopted form of the two parametrisations.

As a concrete illustration of the impact of jet smearing on event reconstruction, we show in figure~\ref{fig:chisqsmear} the distribution of the $\chi^{2,(6)}_\text{min}$ variable in the case of triple Higgs boson production in the SM (thus for a scenario with $c_3=d_4=0$). This composite observable, defined in section~\ref{sec:analysis}, quantifies the compatibility of an event with the hypothesis of the production of three Higgs bosons decaying to six $b$-jets. A shift of the distribution towards larger values is clearly visible when ATLAS-like or CMS-like smearing is applied, as expected from the degradation of jet energy resolution. For less composite observables such as the transverse momenta of the leading $b$-jets, the smearing effects are found smaller and thus not shown.

\begin{table}\renewcommand{\arraystretch}{1.5}
  \centering \begin{tabular}{ccc}
  \multicolumn{3}{c}{\bf {Reducible Backgrounds}}\\
  Process & $\sigma_\mathrm{GEN}$ [fb] & $\sigma_\mathrm{GEN} \times k_\mathrm{fac}  \times \mathcal{P}(6b)$ [fb]\\[-.1cm]
\noalign{\smallskip}\hline\noalign{\smallskip}
$(b\bar{b}) (b\bar{b}) (c\bar{c})$ & $73.0 \times 10^3$ & $762$\\
$(b\bar{b}) (c\bar{c}) (c\bar{c})$ & $72.0\times 10^3 $ & $10.4$ \\
$(c\bar{c}) (c\bar{c}) (c\bar{c})$ & $21.8 \times 10^3$ & $4.36 \times 10^{-2}$ \\
$(b\bar{b}) (b\bar{b}) (jj)$* & $3.09 \times 10^6$ & $161$\\
$(b\bar{b}) (jj) (jj)$* & $1.07 \times 10^8$ & $1.54$ \\
$(c\bar{c}) (c\bar{c}) (jj)$* & $1.73 \times 10^5$ & $3.46 \times 10^{-3}$\\
$(c\bar{c}) (jj) (jj)$* & $ 6.67 \times 10^7$ & $6.67 \times 10^{-3}$\\
$(jj) (jj) (jj)$* & $3.04 \times 10^9$ & $6.08 \times 10^{-3}$\\
\noalign{\smallskip}\hdashline
& sum: & $\sim 935$ \\
  \end{tabular}
  \caption{Reducible background processes considered in the six $b$-jet analysis. The second column shows the generation-level cross sections obtained after applying the cuts of eq.~\eqref{eq:cuts}, while the third column reports the cross sections obtained after incorporating $b$-tagging performance and a $k$-factor of 2. Processes marked with a ``*'' have been generated with \texttt{ALPGEN}, using the MRSTLO** PDF set, and we only include processes that can mimic QCD-induced six-$b$-jet production.}\label{table:reducible}
\end{table}

For $b$-tagging, we assume a ``central'' efficiency of $\mathcal{P}(b \to b) = 0.85$ for correctly identifying a $b$-jet. To estimate the impact of the reducible background contributions, we take misidentification probabilities of $\mathcal{P}(j \to b) = 0.01$ for light jets and $\mathcal{P}(c \to b) = 0.10$ for charm jets. We then compute cross sections for six-jet processes involving charm and/or light jets, applying the same generation-level cuts as for the irreducible backgrounds (see eq.~\eqref{eq:cuts}). The results are presented in table~\ref{table:reducible}, which shows the generation-level cross sections and the effective ones derived after requiring that all six jets are identified as $b$-jets.\footnote{For several processes, cross section estimates were obtained with the \texttt{ALPGEN} Monte Carlo event generator~\cite{Mangano:2002ea} and the MRSTLO** PDF set~\cite{Sherstnev:2008dm}.} Assuming comparable analysis efficiencies as for the signal, we find that reducible backgrounds contribute at the level of $\mathcal{O}(10\%)$. However, as tree-level estimates, these numbers carry relatively large theoretical uncertainties and should ultimately be reassessed in a dedicated phenomenological study once more advanced Monte Carlo tools become available.

\subsection{Probing Triple-Higgs production at Future Colliders}\label{sec:analysis}
We now describe the two complementary analysis strategies explored in this work: a conventional cut-based approach employing ``rectangular’’ selections, and a machine-learning approach based on gradient boosting using \texttt{XGBoost}, an optimised distributed gradient boosting library designed to be highly efficient, flexible and portable~\cite{Chen_2016}. The use of machine-learning algorithms on Monte Carlo samples is well justified in our setup, as the inclusion of four-momentum smearing prevents overtraining on sharp features that would not survive in realistic experimental conditions. 

In both approaches we first impose minimal preselection requirements: events must contain at least six $b$-jet candidates identified by tracing the presence of truth-level $b$-hadrons inside the jets and the tagging performance defined in section~\ref{sec:sim}, and satisfying
\begin{equation}
  p_{T,b} > 35~\mathrm{GeV}\;, \quad |\eta_b| < 3.0 \quad\text{and}\quad \Delta R_{b,b} > 0.3\;.
\end{equation}
In particular, the pseudo-rapidity requirement corresponds to a forward angle of about $5.7^\circ$, which may be achievable at a future collider~\cite{Helsens:2019bfw}. After this preselection, higher-level observables are constructed from the six leading $b$-jets in transverse momentum~\cite{Papaefstathiou:2019ofh, Papaefstathiou:2020lyp, Papaefstathiou:2023uum}. These observables are then used either to define further cuts in the cut-based strategy, or to train the \texttt{XGBoost} classifier to separate signal from background. The training is performed in the Standard Model scenario, with $(c_3,d_4)=(0,0)$, and subsequently applied across the $(c_3,d_4)$ plane to extract signal selection efficiencies within the entire parameter space.

The first higher-level observable that we construct from the six $b$-jets is the $\chi^{2,(6)}$ variable defined as
\begin{equation}
\chi^{2, (6)} = \sum_{qr \in I} (m_{qr}-m_h)^2\;,
\label{eq:chi2}\end{equation}
where $I=\{ b_{i_1}b_{i_2},\, b_{i_3}b_{i_4},\, b_{i_5}b_{i_6}\}$ denotes one of the possible 15 pairings of six $b$-tagged jets. Among all possible assignments, we select the combination that minimises the $\chi^{2,(6)}$ value and that thus consists in our best option for reconstructing the three Higgs bosons $h^1$, $h^2$ and $h^3$ (here ordered in transverse momentum). The selection can be further refined by considering the individual mass differences $\Delta m = |m_{qr}-m_h|$ appearing in eq.~\eqref{eq:chi2} and on which we impose independent cuts, and we also exploit the transverse momenta of the reconstructed Higgs boson candidates $p_T(h^i)$ with $i=1,2,3$. We next apply further constraints directly on the transverse momenta $p_{T,b_i}$ of the three leading $b$-jets (with $i=1,2,3$), this last set of cuts constituting the second most powerful discriminating variable in our list, and we introduce two geometric observables, the angular separation of the $b$-jets within each reconstructed Higgs candidate $\Delta R_{bb}(h^i)$ and the distance between each pair of reconstructed Higgs bosons $\Delta R(h^i,h^j)$. Finally, for the \texttt{XGBoost}-based analysis we additionally include the invariant mass of the system comprising the six leading $b$-jets as an input feature.

\begin{table}\renewcommand{\arraystretch}{1.4}\setlength\tabcolsep{8pt}
\begin{center} \begin{tabular}{ l l }  
  \multicolumn{2}{c}{\bf {Cut Details}}\\
  Observable & Threshold \\ \hline
  $p_{T,b}>$ & $35.0$ GeV\\ 
  $|\eta_{b}|<$ &  $3.0$ \\ 
  $\Delta R_{bb}>$ & $0.3$\\
  \hdashline
  $p_{T,b_i}>$ &$[170.0, 135.0, 35.0]$ GeV\\
  $\chi^{2,(6)} <$  & $26.0$ GeV\\
  $\Delta m_i<$& $[8, 8, 8]$ GeV\\
  $\Delta R_{bb}(h^i)<$ & $[3.5, 3.5, 3.5]$\\
  $\Delta R(h^i,h^j)<$ & $[3.5, 3.5, 3.5]$\\
  $p_{T}(h^i)>$ &$[200.0, 190.0, 20.0]$ GeV\\
  \end{tabular}
  \caption{$b$-jet candidate requirements (upper panel) and rectangular cuts (lower panel) applied in the cut-based analysis, with $i,j=1,2,3$. For the $\Delta R(h^i, h^j)$ selections, the three possible pairings are ordered as $(1,2)$, $(1,3)$ and $(2,3)$ while for any cut depending on an indexed quantity, the thresholds are provided in increasing order of the index from left to right. \label{tab:cuts}} 
  \end{center}
\end{table}

For the cut-based analysis, we then optimise the selection thresholds following the strategy of~\cite{Papaefstathiou:2023uum}, and we summarise our choice of cuts in table~\ref{tab:cuts}. In the \texttt{XGBoost} approach, by contrast, the values of the observables defined above are retained for each event and directly used as training features. The performance of both analyses is next quantified through the corresponding confusion matrices from which we extract the efficiencies: for the signal sample this corresponds to the true positive rate, while for the background samples it corresponds to the false positive rate.

To derive constraints on the $(c_3,d_4)$ plane under the null hypothesis of SM triple-Higgs production, we assume that the event counts for both the signal and the background follow Gaussian statistics. The uncertainty $\delta_\mathrm{SM+B}$ on the total number of expected SM events is then given by
\begin{equation}\label{eq:deltaSMplusB}
    \delta_\mathrm{SM+B} = \sqrt{S_{\mathrm{SM}} + B + (\alpha B)^2}\;,
\end{equation}
where $S_\mathrm{SM}$ is the expected number of SM signal events at integrated luminosity $L$, $B$ is the total background yield at the same luminosity, and $\alpha$ parametrises the systematic uncertainty on the background. This allows us to define, for each parameter point $(c_3,d_4)$, the test statistic
\begin{equation}
  \chi^2(c_3,d_4) = \left[\frac{ S_\mathrm{SM} - S(c_3,d_4)}{ \delta_\mathrm{SM+B}}\right]^2 ,
\end{equation}
where $S(c_3,d_4)$ denotes the expected signal yield for arbitrary $c_3$ and $d_4$ values. When including external information on $c_3$ (like possible constraints emerging from Higgs boson pair production) studies, we modify the above test by adding a Gaussian prior,
\begin{equation}\label{eq:test}
\chi^2_\text{tot}(c_3,d_4) = \chi^2(c_3,d_4) + \left( \frac{c_3}{ \delta c_3} \right)^2 ,
\end{equation}
with $\delta c_3$ being the expected $1\sigma$ precision on $c_3$ (centred at $c_3=0$). In practice we take $\delta c_3 = 0.05$, following table 12 in~\cite{deBlas:2019rxi}.

We first extract 95\% confidence level bounds from triple-Higgs production alone, without assuming any external knowledge of $c_3$, and therefore ignoring the second term in eq.~\eqref{eq:test}. We derive in this case simultaneous constraints on $c_3$ and $d_4$ by enforcing $\Delta \chi^2 = \chi^2(c_3,d_4) - \chi^2_\mathrm{min} < 5.99$, with $\chi^2_\mathrm{min}$ being the global minimum over the considered parameter space. In addition, one-dimensional limits on one of the two new physics parameters are obtained by profiling over the other parameter. For example, the $d_4$ limits are determined by minimising:
\begin{equation}\label{eq:profiling}
    \chi^2_p(d_4) = \min_{c_3} \chi^2(c_3, d_4)\;,
\end{equation}
while bounds on $c_3$ are obtained similarly from $\chi^2_p(c_3)$. This time, the allowed region of parameter space is obtained by requiring $\Delta \chi^2 < 3.84$. As a second exercise, we assess the impact of external information on the trilinear coupling $c_3$ by including the prior in eq.~\eqref{eq:test}. The one-dimensional bounds on $d_4$ are then obtained by profiling as in eq.~\eqref{eq:profiling}, but using the total statistic $\chi^2_\text{tot}(c_3,d_4)$ instead.

\section{Sensitivity to the Higgs Self-Interactions}\label{sec:results}
We now apply the analyses introduced in section~\ref{sec:pheno} to both the signal and the backgrounds, and translate the results into constraints on the trilinear and quartic Higgs self-coupling modifiers $c_3$ and $d_4$. Furthermore, we explicitly assess the effects of detector performance, background normalisation and EFT truncation, and we additionally overlay our findings with constraints originating from perturbative unitarity. Finally, extrapolations for various centre-of-mass energies and integrated luminosities relevant for post-LHC collider scenarios are presented.

\begin{figure}
    \centering
    \includegraphics[width=\linewidth]{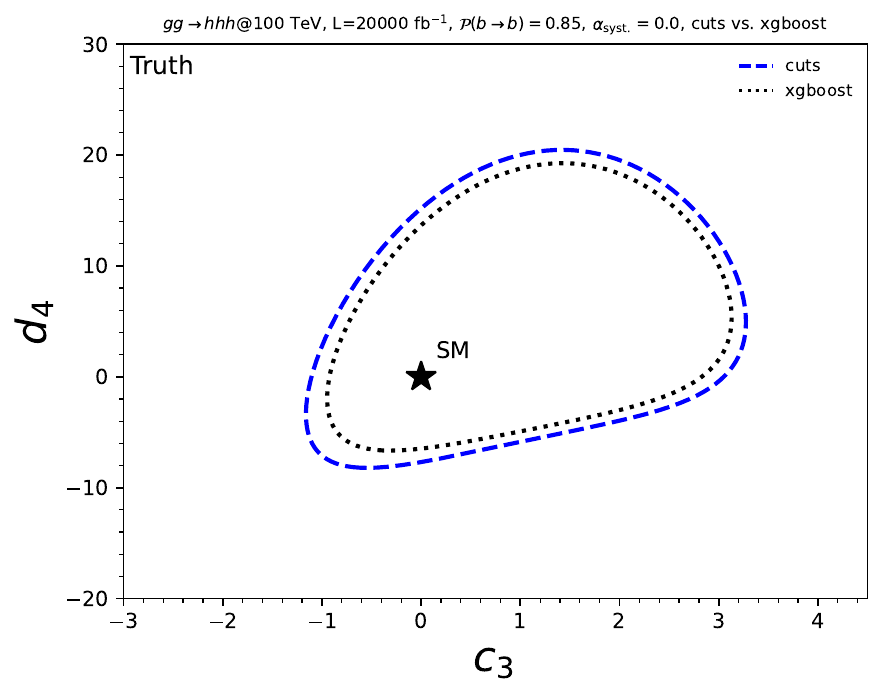}
    \caption{Comparison of the $2\sigma$ limits on the $(c_3,d_4)$ plane expected from $L=20$~ab$^{-1}$ of proton-proton collisions at $\sqrt{s}=100$~TeV, obtained with the cut-based analysis (dashed blue) and with the \texttt{XGBoost} approach (dotted black). Results are shown without smearing and systematics.}
    \label{fig:analysiscomp}
\end{figure}
In figure~\ref{fig:analysiscomp}, we present a comparison between the expected $2\sigma$ constraints on the $(c_3, d_4)$ plane obtained with our cut-based analysis (dashed blue) and those from the gradient-boosting approach implemented via \texttt{XGBoost} (dotted black), ignoring for the moment the systematics (\textit{i.e.}\ $\alpha=0$ in eq.~\eqref{eq:deltaSMplusB}), smearing effects and any truncation of the signal matrix element, and considering 20~ab$^{-1}$ of proton-proton collisions at $\sqrt{s}=100$~TeV. Both strategies consistently indicate that values of $c_3$ outside the approximate interval $[-1,3]$ and of $d_4$ outside $[-10,20]$ would be excluded at the $2\sigma$ level, illustrating the reach of a 100~TeV collider in probing departures from the SM Higgs self-couplings. In particular, the \texttt{XGBoost} strategy is found to yield an only moderately stronger sensitivity than the conventional cut-based method. However, it has the additional advantage of retaining a larger number of events for both signal and background samples, as is quantitatively reflected in the event yields of table~\ref{table:processes}. Although the relative nature of the constraints with respect to the SM point somewhat limits the apparent improvement, the gain in statistical significance is striking: while the cut-based analysis achieves a significance of about 1 for the $c_3=d_4=0$ scenario, the \texttt{XGBoost} analysis nearly doubles this value to 2. This highlights the potential of multivariate techniques to substantially enhance sensitivity in challenging final states.

\begin{figure*}
    \centering
    \includegraphics[width=0.48\linewidth]{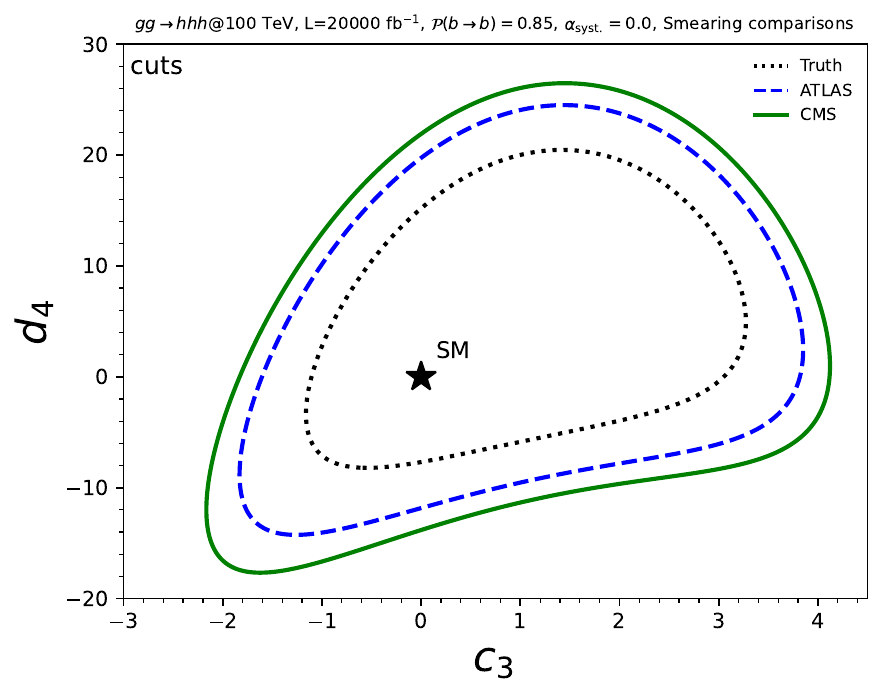}\hfill 
    \includegraphics[width=0.50\linewidth]{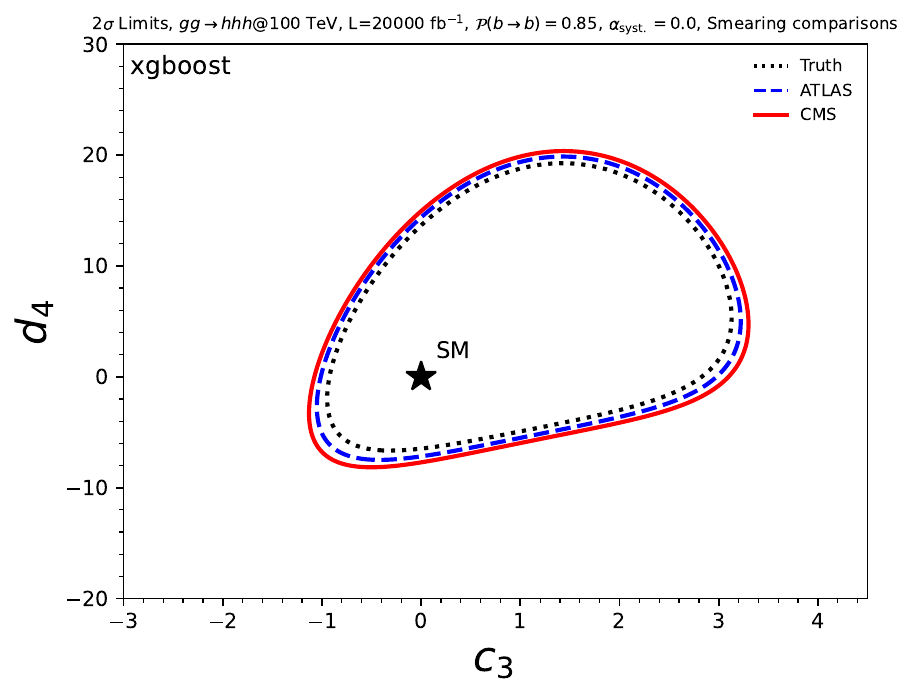}
    \caption{Comparison of the $2\sigma$ limits on the $(c_3,d_4)$ plane expected from $L=20$~ab$^{-1}$ of proton-proton collisions at $\sqrt{s}=100$~TeV, obtained with the cut-based analysis (left) and the \texttt{XGBoost} analysis (right). Results are shown for the Monte Carlo truth (dotted black, no smearing) as well as for the considered ATLAS-like (dashed blue) and CMS-like (solid red) jet smearing models.}
    \label{fig:smearingscomp}
\end{figure*}

\begin{table}\renewcommand{\arraystretch}{1.25}
  \begin{center}\resizebox{\linewidth}{!}{\begin{tabular}{ l l l } 
  Analysis & \multicolumn{2}{c}{\bf {No $c_3$ prior}}\\
  MC truth - Cuts                   & $c_3 \in [-1.0,  3.1]$ & $d_4 \in [-6.9, 20]$ \\ 
  MC truth - \texttt{XGBoost}       & $c_3 \in [-0.82, 3.0]$ & $d_4 \in [-5.6, 19]$ \\
  ATLAS smearing - \texttt{XGBoost} & $c_3 \in [-0.90, 3.1]$ & $d_4 \in [-6.3, 19]$ \\
  CMS smearing - \texttt{XGBoost}   & $c_3 \in [-0.97, 3.2]$ & $d_4 \in [-6.9 , 19]$ \\[.1cm]
  \hline
  Analysis & {\bf {Gaussian $c_3$ prior}}\\
  MC truth - Cuts:                  & $d_4 \in [-6.7, 14]$ \\ 
  MC truth - \texttt{XGBoost}       & $d_4 \in [-5.6, 13]$ \\
  ATLAS smearing - \texttt{XGBoost} & $d_4 \in [-6.2, 13]$ \\
  CMS smearing - \texttt{XGBoost}   & $d_4 \in [-6.7, 14]$ \\
  \end{tabular}}
  \caption{Expected $2\sigma$ constraints on the self-coupling modifiers $c_3$ and $d_4$ obtained with the cut-based and \texttt{XGBoost} analysis methods at $\sqrt{s}=100$~TeV and for a luminosity of $L=20$~ab$^{-1}$. We ignore the truncation of the EFT expansion and the systematics, and compute the sensitivity using the Monte Carlo (MC) truth as well as after CMS-like and ATLAS-like smearing. The sensitivity to one parameter is reported after profiling over the other parameter without (upper panel) or after including  (lower panel) a Gaussian prior on $c_3$ with width $\delta c_3=0.05$.}
\label{tab:1Dconstraints}
\end{center}
\end{table}

In figure~\ref{fig:smearingscomp}, we compare the $2\sigma$ limits obtained under different detector resolution choices, considering both the cut-based analysis (left panel) and the \texttt{XGBoost} analysis (right panel). The limits derived with ATLAS-like and CMS-like smearing models, as described in section~\ref{sec:sim}, are shown together with the Monte Carlo truth results of figure~\ref{fig:analysiscomp}. As expected, the inclusion of smearing slightly weakens the bounds relative to the idealised truth-level case, but the effect remains modest and the overall sensitivity stays within the same ballpark. We find that the dependence on the smearing model is more pronounced in the cut-based analysis, where ATLAS-like and CMS-like resolutions yield visibly different contours featuring bounds that are much more apparently weaker than in the idealised case of the Monte Carlo truth. By contrast, in the \texttt{XGBoost} approach the limits obtained with the two smearing models are nearly indistinguishable. This stability arises because the \texttt{XGBoost} classifier is retrained for each detector model, thereby absorbing part of the resolution effects. Consequently, smearing effects are essentially negligible in the multivariate analysis, highlighting another advantage of machine-learning approaches over traditional cut-based strategies. As an alternative to present the results, we display in table~\ref{tab:1Dconstraints} the corresponding one-dimensional limits on $c_3$ and $d_4$, as obtained after respectively marginalising over $d_4$ and $c_3$, following eq.~\eqref{eq:profiling}. We both consider a situation in which no prior is imposed on the $c_3$ parameter (upper panel of the table) and after assuming that $c_3$ will be obtained at 5\% level and with a value compatible with the SM expectation. In the remainder of this work, we consequently adopt as our central benchmark the limits obtained with the \texttt{XGBoost} analysis under the CMS-like smearing, with no additional modelling of systematic uncertainties (\textit{i.e.}\ $\alpha=0$ in eq.~\eqref{eq:deltaSMplusB}), at $\sqrt{s}=100$~TeV and an integrated luminosity of $L=20$~ab$^{-1}$.

\begin{figure}
    \centering
    \includegraphics[width=\linewidth]{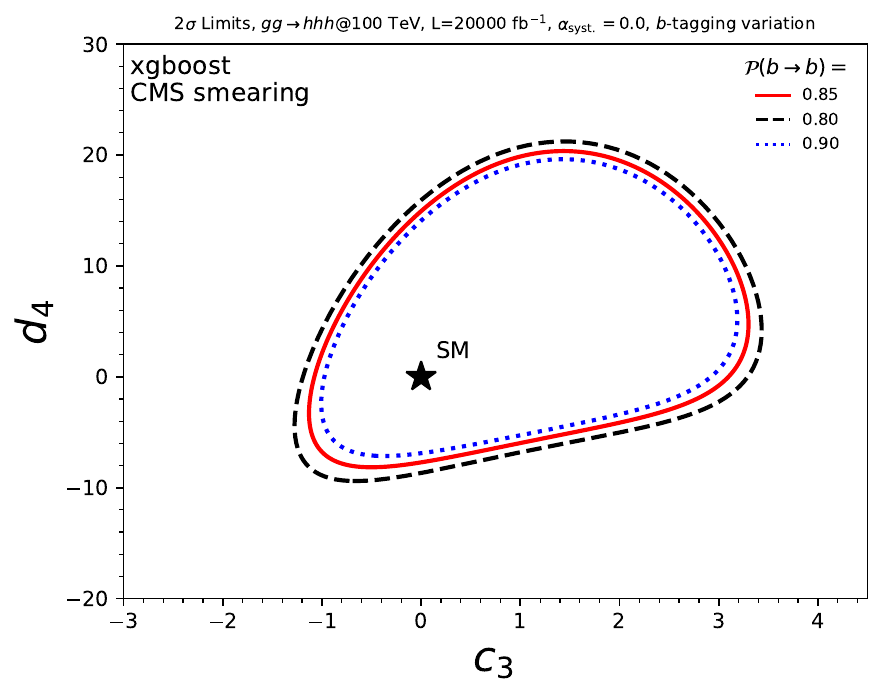}
    \caption{Impact of the $b$-tagging efficiency on the $2\sigma$ reach in the $(c_3,d_4)$ plane, as expected from 20~ab$^{-1}$ of proton-proton collisions at $\sqrt{s}=100$~TeV. Contours are shown for CMS-like jet smearing functions within the \texttt{XGBoost} analysis, and correspond to $\mathcal{P}(b\!\rightarrow\! b)=80\%$ (dashed black), 85\% (solid red) and 90\% (dotted blue).}
    \label{fig:btagcomp}
\end{figure}
To assess the robustness of our results against detector-level effects, we have also studied the impact of varying the $b$-tagging identification efficiency on the extracted limits. In particular, we have considered efficiencies of $\mathcal{P}(b\rightarrow b)=0.80$, 0.85 and 0.90, and we present the corresponding $2\sigma$ contours in figure~\ref{fig:btagcomp}. The differences between the contours are minimal, demonstrating that $\mathcal{O}(10\%)$ variations in the $b$-tagging efficiency have only a marginal impact on the reach. These results indicate that the extraction of constraints on the Higgs self-couplings at a future 100~TeV collider is not expected to be strongly limited by realistic uncertainties in $b$-tagging performance, which further underlines the robustness of the multivariate approach.

%This insensitivity is a notable feature of the multivariate strategy: since the \texttt{XGBoost} classifier is retrained for each configuration, it effectively absorbs such variations and maintains a stable performance

\begin{figure}
    \centering
    \includegraphics[width=\linewidth]{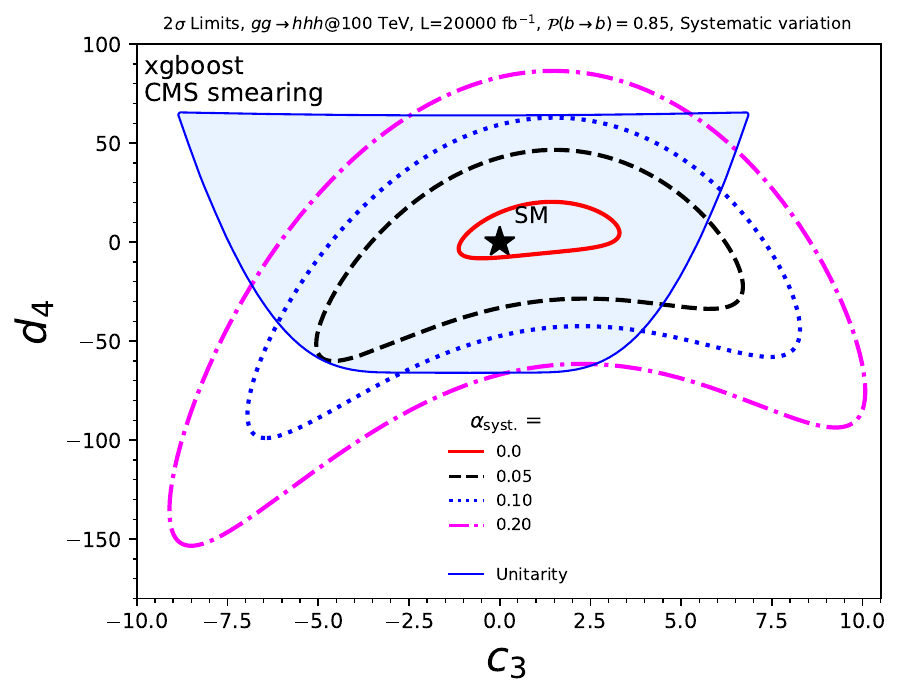}
    \caption{Impact of systematic uncertainties on the expected $2\sigma$ reach in the $(c_3,d_4)$ plane from 20~ab$^{-1}$ of proton-proton collisions at $\sqrt{s}=100$~TeV. Results are shown for CMS-like jet smearing functions within the \texttt{XGBoost} analysis. The contours correspond to $\alpha=5\%$ (dashed black), $10\%$ (dotted blue) and $20\%$ (dash-dotted magenta), and they are compared to the ideal case with $\alpha=0$ (solid red). The parameter space region allowed by tree-level perturbative unitarity is also indicated (shaded blue).}
    \label{fig:systcomp}
\end{figure}

\begin{table}\renewcommand{\arraystretch}{1.25}
  \begin{center}\begin{tabular}{ c l l} 
    $\alpha$ & \multicolumn{2}{c}{\bf No $c_3$ prior}\\
    $0.00$ & $c_3 \in [-0.97, 3.2]$ & $d_4 \in [-6.9 , 19]$\\
    $0.05$ & $c_3 \in [-4.5, 6.3]$  & $d_4 \in [-51,  43]$\\
    $0.10$ & $c_3 \in [-6.3, 7.7]$  & $d_4 \in [-85,  57]$\\
    $0.20$ & $c_3 \in [-8.4, 9.5]$  & $d_4 \in [-130, 78]$\\[.1cm]
    \hline
    $\alpha$ & {\bf {Gaussian $c_3$ prior}}\\
    $0.00$ & $d_4 \in [-6.7, 14]$\\
    $0.05$ & $d_4 \in [-30, 39]$ \\
    $0.10$ & $d_4 \in [-42, 53]$ \\
    $0.20$ & $d_4 \in [-60, 75]$ \\
  \end{tabular}
  \caption{Expected $2\sigma$ constraints on the self-coupling modifiers $c_3$ and $d_4$ for different level of systematics modelled through the parameter $\alpha$ in eq.~\eqref{eq:deltaSMplusB}, as obtained with the \texttt{XGBoost} analysis at $\sqrt{s}=100$~TeV, $L=20$~ab$^{-1}$ and using CMS-like smearing. The results on $c_3$ and $d_4$ are reported after profiling over the other parameter and either without (upper panel) or after including (lower panel) a Gaussian prior on $c_3$ with width $\delta c_3=0.05$. \label{tab:syst}}
\end{center}
\end{table}

Systematic uncertainties represent a key ingredient in assessing the robustness of the projected limits. We model their impact through the parameter $\alpha$ in eq.~\eqref{eq:deltaSMplusB}, and evaluate the $2\sigma$ sensitivity for $\alpha=5\%$, $10\%$ and $20\%$. The corresponding contours are shown in figure~\ref{fig:systcomp}. For reference, we also include the idealised case with $\alpha=0$, together with the region allowed by tree-level perturbative unitarity determined as described in section~\ref{sec:theory}. The results demonstrate that even moderate systematic effects degrade the constraints on $c_3$ and $d_4$ by factors of a few. Achieving uncertainties at the level of $\alpha\lesssim 5\%$ will therefore be essential to fully exploit the physics potential of future proton-proton colliders and probe somehow deeply the theoretically-motivated region of the parameter space. Nonetheless, even in the more pessimistic case with $\alpha=20\%$, the analysis remains sensitive to the region consistent with perturbative unitarity in which large new physics effects on triple Higgs production are expected, via either a large value of $c_3$ or of $d_4$, or both. These findings are further illustrated in table~\ref{tab:syst} where we display predictions for the sensitivity to one of the Higgs self-coupling modifiers after profiling over the other, without (upper panel) or with (lower panel) a Gaussian prior on $c_3$. Our results underline thus both the importance of controlling the systematics, and the advantage of multivariate strategies to optimise the discovery potential of anomalous effects.

\begin{figure}
    \centering
    \includegraphics[width=\linewidth]{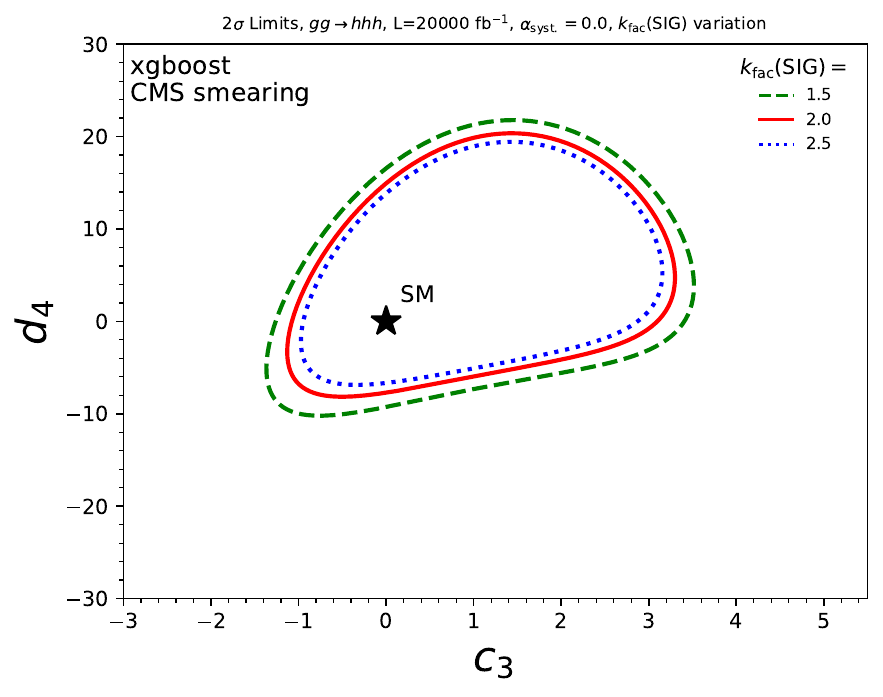}
    \caption{Impact of signal normalisation on the expected $2\sigma$ limits in the $(c_3,d_4)$ plane, for  20~ab$^{-1}$ of proton-proton collisions at $\sqrt{s}=100$~TeV. Results are shown for the \texttt{XGBoost} analysis with CMS-like jet smearing, varying the uniform $k$-factor applied to all background processes. We consider $k_\mathrm{fac}(\mathrm{SIG})=1.5$ (dashed green), 2 (solid red) and 2.5 (dotted blue).}
    \label{fig:kfacsigvar}
\end{figure}

\begin{figure}
    \centering
    \includegraphics[width=\linewidth]{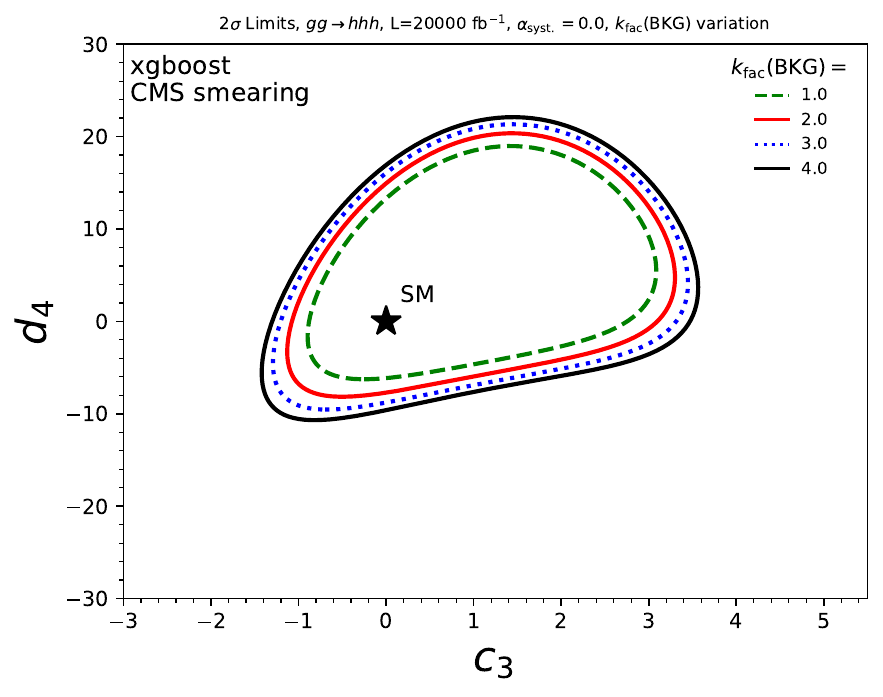}
    \caption{Impact of background normalisation on the expected $2\sigma$ limits in the $(c_3,d_4)$ plane, for  20~ab$^{-1}$ of proton-proton collisions at $\sqrt{s}=100$~TeV. Results are shown for the \texttt{XGBoost} analysis with CMS-like jet smearing, varying the uniform $k$-factor applied to all background processes. We consider $k_\mathrm{fac}(\mathrm{BKG})=1$ (dashed green), 2 (solid red), 3 (dotted blue) and 4 (solid black).}
    \label{fig:kfacbkgvar}
\end{figure}

\begin{table}\renewcommand{\arraystretch}{1.25}
  \begin{center}\begin{tabular}{ c l l} 
    $k_\mathrm{fac}(\mathrm{SIG})$ & \multicolumn{2}{c}{\textbf{No $c_3$ prior}}\\
    1.5 & $c_3 \in [-1.2, 3.3]$ & $d_4 \in [-8.6, 21]$ \\  
    2.0 & $c_3 \in [-1.0,  3.2]$ & $d_4 \in [-6.9, 19]$\\
    2.5 & $c_3 \in [-0.84, 3.0]$ & $d_4 \in [-5.8, 19]$ \\ 
    \hline
    $k_\mathrm{fac}(\mathrm{SIG})$ & \textbf{Gaussian $c_3$ prior}\\
    1.5 & $d_4 \in [-8.0, 15]$ \\    
    2.0 & $d_4 \in [-6.7, 14]$ \\
    2.5 & $d_4 \in [-5.7, 13]$ \\ 
  \end{tabular}
  \caption{Expected $2\sigma$ constraints on the self-coupling modifiers $c_3$ and $d_4$ for different signal $k$-factors, obtained with the \texttt{XGBoost} analysis at $\sqrt{s}=100$~TeV, $L=20$~ab$^{-1}$ and using CMS-like smearing. Systematic uncertainties are neglected ($\alpha=0$ in eq.~\eqref{eq:deltaSMplusB}), and the results are reported after profiling over the other parameter without (upper panel) or with (lower panel) a Gaussian prior on $c_3$ with width $\delta c_3=0.05$ (lower panel). \label{tab:constraintskfacsig}}
\end{center}
\end{table}

\begin{table}\renewcommand{\arraystretch}{1.25}
  \begin{center}\begin{tabular}{ c l l} 
    $k_\mathrm{fac}(\mathrm{BKG})$ & \multicolumn{2}{c}{\textbf{No $c_3$ prior}}\\
    1.0 & $c_3 \in [-0.77, 3.0]$ & $d_4 \in [-5.3, 18]$ \\  
    2.0 & $c_3 \in [-1.0,  3.2]$ & $d_4 \in [-6.9, 19]$\\
    3.0 & $c_3 \in [-1.1,  3.3]$ & $d_4 \in [-8.0, 20]$ \\ 
    4.0 & $c_3 \in [-1.2,  3.4]$ & $d_4 \in [-9.0, 21]$ \\[.1cm]
    \hline
    $k_\mathrm{fac}(\mathrm{BKG})$ & \textbf{Gaussian $c_3$ prior}\\
    1.0 & $d_4 \in [-5.3, 12.5]$ \\    
    2.0 & $d_4 \in [-6.7, 14]$ \\
    3.0 & $d_4 \in [-7.6, 15]$ \\ 
    4.0 & $d_4 \in [-8.3, 16]$ \\
  \end{tabular}
  \caption{Expected $2\sigma$ constraints on the self-coupling modifiers $c_3$ and $d_4$ for different uniform background $k$-factors, obtained with the \texttt{XGBoost} analysis at $\sqrt{s}=100$~TeV, $L=20$~ab$^{-1}$ and using CMS-like smearing. Systematic uncertainties are neglected ($\alpha=0$ in eq.~\eqref{eq:deltaSMplusB}), and the results are reported after profiling over the other parameter without (upper panel) or with (lower panel) a Gaussian prior on $c_3$ with width $\delta c_3=0.05$ (lower panel). \label{tab:constraintskfac}}
\end{center}
\end{table}

Up to this point the signal, and the dominant backgrounds, most notably the QCD $6b$-jet process, have been modelled at tree level with the associated production cross section supplemented by a flat $k$-factor of 2 to account for higher-order corrections. Since this approximation introduces an intrinsic uncertainty, it is important to evaluate how variations in the signal and background normalisation affect the extracted limits on the Higgs self-coupling modifiers. To this end, we separately vary the signal $k$-factor within $k_\mathrm{fac}(\mathrm{SIG}) \in [1.5, 2, 2.5]$ and the common background $k$-factor within $k_\mathrm{fac}(\mathrm{BKG}) \in [1, 2, 3, 4]$, and then derive the corresponding $2\sigma$ constraints. The results, obtained with the \texttt{XGBoost} analysis and CMS-like smearing, are shown in figures~\ref{fig:kfacsigvar} and~\ref{fig:kfacbkgvar}, for the signal and background variations, respectively. We collect the corresponding one-dimensional numerical intervals obtained from eq.~\eqref{eq:profiling} in tables~\ref{tab:constraintskfacsig} and~\ref{tab:constraintskfac}. In all cases, the impact of signal and background normalisation is remarkably mild: even for a fourfold enhancement of the background cross sections, the deterioration of the sensitivity remains limited. This robustness indicates that our projections are not strongly driven by the precise modelling of higher-order corrections on the signal or background, and further suggests that the omission of challenging multi-jet reducible backgrounds is unlikely to qualitatively alter the conclusions. 

\begin{table}\renewcommand{\arraystretch}{1.35}\setlength\tabcolsep{15pt}
 \begin{center}
\begin{tabular}{ l l } 
 Truncation scheme & Bounds\\ \hline
 Quadratic & $d_4 \in [-3.3 , 7.1]$ \\
 Cubic & $d_4 \in [-3.5 , 7.1]$ \\
 No truncation & $d_4 \in [-6.7 , 14]$ \\
  \end{tabular}
  \caption{Expected $2\sigma$ constraints on the quartic self-coupling modifier $d_4$ under different truncation scenarios of the squared matrix-element expansion. To avoid unphysical (negative) cross sections, constraints for the linear truncation scheme are not quoted and we impose a Gaussian prior on $c_3$ to confine the analysis to the physically viable region around $c_3\simeq 0$. All results correspond to $\sqrt{s}=100$~TeV and an integrated luminosity of $L=20$~ab$^{-1}$, as well as no systematics ($\alpha\!=\!0$ in eq.~\eqref{eq:deltaSMplusB}).}
\label{tab:constraintstruncation}
\end{center}
\end{table}

We now turn to the theoretical uncertainties associated with the different possible truncation schemes for the matrix-element expansion, as discussed in section~\ref{sec:theory}. First, our (not shown) results confirm that the contours obtained under a linear truncation scheme are not physically meaningful, since the corresponding predictions for the sensitivity at $\sqrt{s}=100$~TeV extend deep into regions where the triple-Higgs cross section becomes negative. For this reason, we restrict our presentation in table~\ref{tab:constraintstruncation} to the quadratic (first line), cubic (second line) and untruncated (last line) cases, and we always impose the Gaussian prior on $c_3$ assuming an SM-like trilinear Higgs coupling. This effectively confines the fit to the vicinity of $c_3\simeq 0$, where the cross section remains positive and well behaved. The comparison of the obtained bounds reveals that quadratic and cubic truncations yield tighter and similar constraints on the quartic coupling modifier $d_4$ than the untruncated case. The almost identical results for the quadratic and cubic truncation schemes indicate that the dominant sensitivity arises from the quadratic terms in the expansion, while the cubic contributions play only a subleading role. In contrast, retaining the quartic term without truncation broadens the allowed range, hence leading to weaker bounds. Nevertheless, the overall impact of the truncation scheme is moderate: the order of magnitude of the limits on $d_4$ remains stable, with the allowed range varying by at most a factor of two across all consistent scenarios.

\begin{figure*}
  \centering
  \includegraphics[width=0.485\linewidth]{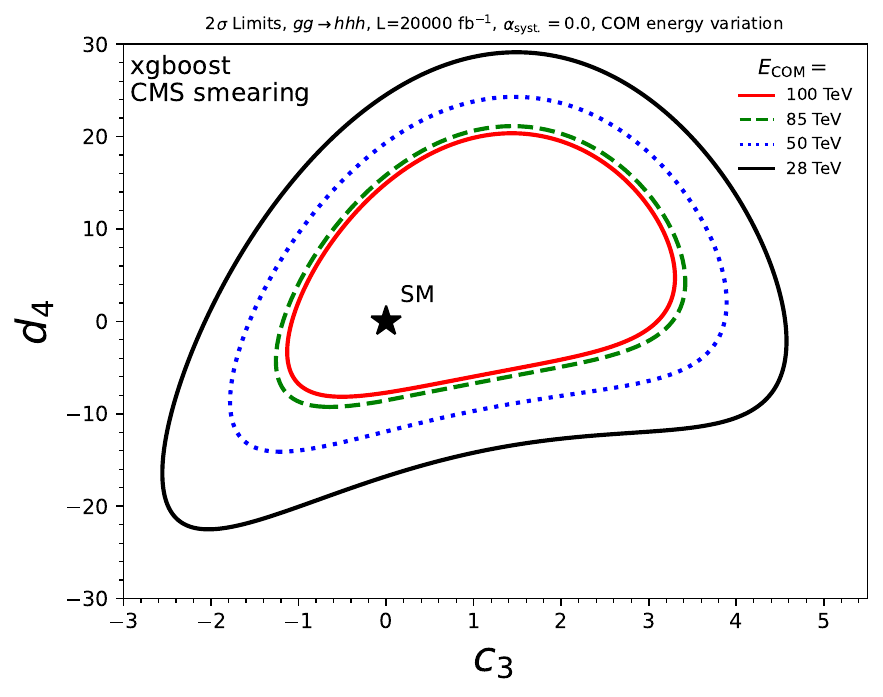} \hfill
  \includegraphics[width=0.485\linewidth]{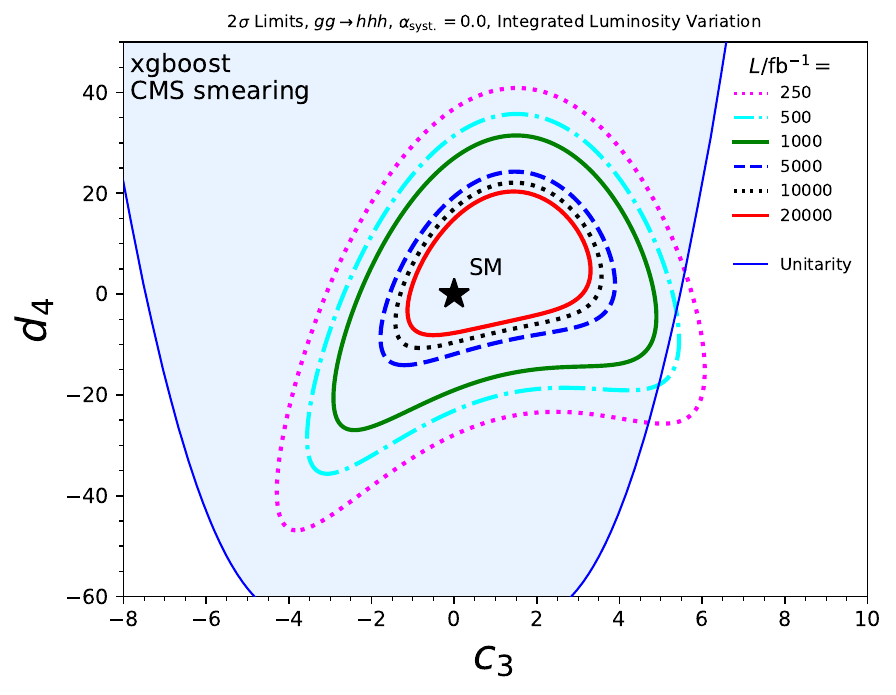}
  \caption{Comparison of the expected $2\sigma$ constraints on the $(c_3, d_4)$ plane obtained with the \texttt{XGBoost} analysis using CMS-like smearing, extrapolated to different proton-proton centre-of-mass energies assuming an integrated luminosity of $20$~ab$^{-1}$ (left) and to different integrated luminosities assuming $\sqrt{s}=100$~TeV (right). The region allowed by perturbative unitarity is indicated through the shaded blue area (right panel).} \label{fig:FutureCollider}
\end{figure*}

Finally, we consider variations in the centre-of-mass energy $\sqrt{s}$ and integrated luminosity $L$, thus probing several of the regimes proposed for a future post-LHC proton-proton collider~\cite{FCC:2025uan}. In the left panel of figure~\ref{fig:FutureCollider}, we display results for $L=20$~ab$^{-1}$ and $\sqrt{s}= 28$~TeV (solid black), 50~TeV (dotted blue), 85~TeV (dashed green) and 100~TeV (solid red), whereas in the right panel of the figure, we investigate a setup in which $\sqrt{s}=100$~TeV and $L=250$~fb$^{-1}$ (dashed pink), 500~fb$^{-1}$ (solid turquoise), 1~ab$^{-1}$ (solid green), 5~ab$^{-1}$ (solid blue), 10~ab$^{-1}$ (dashed black) and 20~ab$^{-1}$ (solid red). To assess the dependence of the sensitivity on the centre-of-mass energy, we have extrapolated the results obtained with the \texttt{XGBoost} analysis and CMS-like smearing by rescaling all cross sections by $s/(100~\mathrm{TeV})^2$, which approximates the change expected in the effective gluon-gluon luminosity. The resulting contours in the $(c_3, d_4)$ plane show that the constraints are only mildly affected when lowering the centre-of-mass energy from 100~TeV to 85~TeV, while at 50~TeV and 28~TeV the reduced signal cross section leads to a deterioration of the bounds by a factor of approximately 1.5 on both $c_3$ and $d_4$. Alternatively, we present the corresponding one-dimensional $2\sigma$ limits on $c_3$ and $d_4$ after marginalising over the other parameter in table~\ref{tab:constraintsenergy}. The results displayed in the right panel of figure~\ref{fig:FutureCollider} show that at lower luminosities like $250$~fb$^{-1}$, the constraints deteriorate by about a factor of 2 compared to the $20$~ab$^{-1}$ benchmark scenario. Nevertheless, a luminosity of at least 1~ab$^{-1}$ is sufficient to guarantee the exploitation of the full potential of future high-energy hadron colliders to probe the different Higgs self-couplings. We however emphasise that  even in the more limited cases, the analysis can still probe parts of the physically motivated region of parameter space in which perturbative unitarity holds (shown through a shaded blue area).

\begin{table}\renewcommand{\arraystretch}{1.25}
  \begin{center}\begin{tabular}{c l l} 
   $\sqrt{s}$ & \multicolumn{2}{c}{\bf {No $c_3$ prior}}\\
   28 TeV  & $c_3 \in [-2.2, 4.3]$  &$d_4 \in [-19 , 27]$ \\  
   50 TeV  & $c_3 \in [-1.5, 3.7]$  &$d_4 \in [-12 , 23]$ \\ 
   85 TeV  & $c_3 \in [-1.1, 3.3]$  &$d_4 \in [-7.8 , 20]$ \\
   100 TeV & $c_3 \in [-0.97, 3.2]$ &$d_4 \in [-6.9 , 19]$\\[.1cm]
  \hline  
  $\sqrt{s}$  & \bf {Gaussian $c_3$ prior}\\
   28 TeV & $d_4 \in [-15 , 22]$ \\    
   50 TeV & $d_4 \in [-10 , 18]$ \\ 
   85 TeV & $d_4 \in [-7.4 , 15]$ \\
   100 TeV & $d_4 \in [-6.7 , 14]$ \\
  \end{tabular}
  \caption{Expected $2\sigma$ constraints on the self-coupling modifiers $c_3$ and $d_4$ for different centre-of-mass energies and an integrated luminosity of 20~ab$^{-1}$, obtained with the \texttt{XGBoost} analysis using CMS-like smearing. Systematic uncertainties are neglected ($\alpha=0$ in eq.~\eqref{eq:deltaSMplusB}), we do not implement any EFT-truncation of the matrix-element expansion, and the sensitivity to one parameter is reported after profiling over the other parameter without (upper panel) or after including (lower panel) a Gaussian prior on $c_3$ with width $\delta c_3=0.05$.}\label{tab:constraintsenergy}
  \end{center}
\end{table}

\section{Summary and Outlook}\label{sec:conclusions}
\begin{figure}
    \centering
    \includegraphics[width=0.85\linewidth]{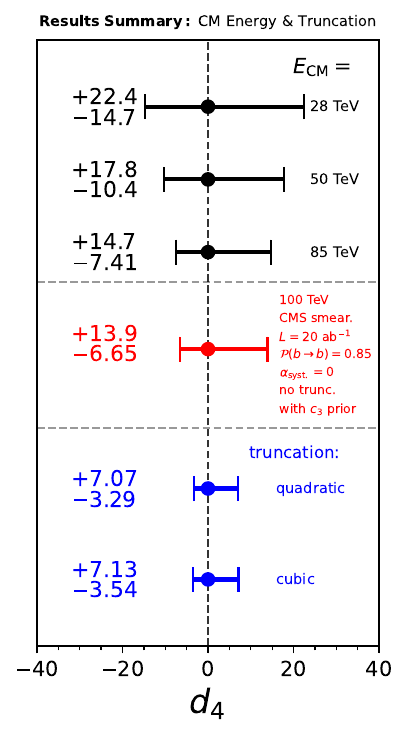}
    \includegraphics[width=0.75\linewidth]{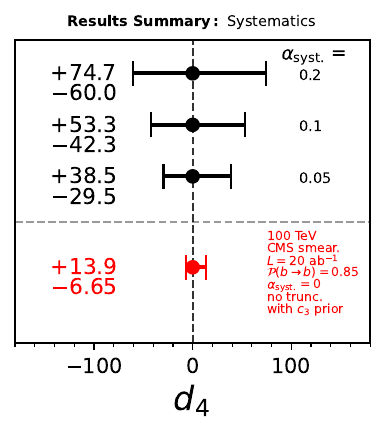}
    \caption{Summary plot of the $2\sigma$ limits on the Higgs quartic self-coupling modifier, for the `central' case considered (in red) with CMS-like smearing, an integrated luminosity of $20$~ab$^{-1}$, and including a Gaussian prior on the trilinear Higgs coupling (with width $\delta c_3=0.05$). Variations of the centre-of-mass energy are shown, together with the impact of the scheme employed to truncate the EFT expansion at the cross section level, and the size of the systematic uncertainties.}
    \label{fig:summary}
\end{figure}

Throughout this study of triple Higgs boson production in the six-$b$-jet final state at future hadron colliders, we have systematically investigated the sensitivity to new-physics effects on the Higgs trilinear and quartic self-couplings via their modifiers $c_3$ and $d_4$. 

Focusing on 20~ab$^{-1}$ of proton-proton collisions at a centre-of-mass energy of 100~TeV, we compared a traditional cut-based strategy with a multivariate approach under both ATLAS-like and CMS-like detector smearing assumptions. The multivariate method was shown to provide stronger sensitivity, while preserving a larger number of selected events, thereby enabling more reliable statistical inference. We next explored a variety of experimental and theoretical systematics. First, we found that achieving less than about approximately $5\%$ of systematic uncertainties will be crucial to optimally probe the perturbative unitarity-allowed region, though even larger uncertainties still permit meaningful constraints. Furthermore, variations in the signal and background normalisation revealed only a mild impact, suggesting that missing higher-order corrections to the signal or reducible backgrounds are unlikely to qualitatively alter our conclusions. In addition, we studied the effect of different EFT truncation choices, showing that quadratic and cubic truncations tighten the $d_4$ bounds relative to the untruncated case, while the overall order of magnitude of the sensitivity to both $c_3$ and $d_4$ remains stable. Finally, extrapolations to different collider energies and luminosities showed that the sensitivity degrades only moderately when moving from 100~TeV to lower centre-of-mass energies, whereas integrated luminosities of at least $\mathcal{O}(1)$~ab$^{-1}$ are required to fully exploit the potential of these machines. In particular, we demonstrated that a 85~TeV proton-proton collider, such as the one envisioned in a reasonably close future, will result in constraints similar to those expected from the higher centre-of-mass energy of 100~TeV previously considered. 

Taken together, our investigations demonstrate that the extraction of meaningful and complementary information on the Higgs self-couplings at future hadron colliders is feasible under a broad range of assumptions. Figure~\ref{fig:summary} collects our results on the $d_4$ sensitivity and its dependence on the above effects, assuming a trilinear coupling consistent with the SM within $5\%$. 

Looking ahead, an important next step will be to combine the six-$b$ channel studied here with other triple-Higgs signatures known to yield constraints of similar order, such as the $4b2\gamma$~\cite{Papaefstathiou:2015paa, Chen:2015gva, Fuks:2015hna, Agrawal:2019bpm, Stylianou:2023tgg}, $4b2\tau$~\cite{Fuks:2015hna, Fuks:2017zkg, Stylianou:2023tgg} and $2b4\tau$~\cite{Dong:2025lkm} final states. Combined with more realistic detector simulations and advanced machine-learning techniques, such global analyses may ultimately demonstrate the possibility of reaching $\mathcal{O}(1)$ precision on the quartic Higgs self-coupling at future hadron colliders.

\section*{Acknowledgments}
AP acknowledges support by the National Science Foundation under Grant No.\ PHY 2210161 and the US Department of Energy, Office of Science, Office of Nuclear Physics under Award Number DE-SC0025728. BF has been supported by Grant ANR-21-CE31-0013 from the \emph{Agence Nationale de la Recherche} (France). This project has received support
from the European Union’s Horizon 2020 research and innovation programme under the
Marie Sklodowska-Curie grant agreement No 945422 [G.T-X.]. This research has been
supported by the Deutsche Forschungsgemeinschaft (DFG, German Research Foundation)
under grant 396021762 - TRR 257 “Particle Physics Phenomenology after the Higgs Dis-
covery”.

\bibliographystyle{JHEP}
\bibliography{biblio.bib}

\end{document}